# Image Transfer through Two Sequential Four-Wave Mixing in a Hot Atomic Vapor

Dong-Sheng Ding, Zhi-Yuan Zhou, Bao-Sen Shi*, Xu-Bo Zou*, and Guang-Can Guo

*Key Laboratory of Quantum Information, University of Science and Technology of China, Hefei 230026, China*

*corresponding author: drshi@ustc.edu.cn

xbz@ustc.edu.cn

Abstract

Efficient wavelength conversion of images has many potential applications in optical communication, sensing, imaging, and quantum information fields. In this work, we report on here the first demonstration of an image transfer between the light of wavelength 780 nm and the light of wavelength 1530 nm by performing two sequential four-wave mixing processes in two different hot atomic rubidium vapor cells. Furthermore, we confirm the persistence of coherence of the input light during this sequential process experimentally. Our results may be useful to the research fields mentioned above.

Efficient frequency conversion of images is an interesting subject of research because of its potential applications in many fields, such as optical communication, imaging, sensing, quantum optics and information, etc. For example, the efficient frequency down-conversion of images has been used to transfer an image imprinted on a pump beam to an entangled photon pair [1-3] for fundamental quantum optical research. With efficient frequency up-conversion of images, we could solve many serious problems that the infrared detection system faces: the limited spectrum response and spatial resolution or sensitivity, the requirement of cooling, and very expensive price, by detecting the up-converted images in the visible region with a very cheap, highly sensitive and low-noise detector with no cryogenic cooling. Such advantages make the up-conversion technology attractive in many fields [4-9]. As a result, there is a strong demand for practical schemes that can efficiently convert frequency of images.

The optical frequency conversion is usually realized by sum frequency or different frequency generation in a nonlinear crystal, but relatively low conversion efficiency makes a laser with high intensity or a resonant cavity been a prerequisite for efficient conversion [10-12]. A promising way for solving the above problem is realizing the frequency conversion in an atomic ensemble with a diamond- or a ladder-type configuration [13-17]. However, these experiments are achieved without images. So far, there is no any experimental report about the coherent frequency conversion of images.

In this work, using a ladder-type configuration of the rubidium (Rb) atom and taking a light with orbital angular momentum (OAM) as an image, (A light with OAM, regarded as a kind of image, has many exciting applications in optical communications [18-20], quantum information field [21], etc.) we demonstrate experimentally that an image imprinted on the light of 780 nm, corresponding to the D2 line of Rb atomic transition, can be transferred to the light of 1530 nm, which is in the transmission window of an optical fibre, with four-wave mixing (FWM) in a hot Rb vapor cell at the first. We then convert this light with transferred image back into the light of wavelength 780 nm in a reverse procedure in another hot Rb vapor cell. Furthermore, we experimentally confirm the persistence of coherence of the input light during this sequential process.

The results make promising progress by demonstrating for the first time the coherent image imprinted on light transfer in and out of the telecom window by FWM in a hot atomic ensemble. Besides, our scheme is very simple compared with the schemes with a nonlinear crystal [10-12]: neither a resonant cavity nor a high power laser is required. Another big advantage is that the setup can be minimized. In addition to many potential applications in traditional optical communication, imaging, night-vision, our results may be used in long-distance quantum communication using atomic-based memories [22-24]. This experiment mimics two important steps along this direction: image imprinted on light transfer in and out of the telecom window.

Taking into account the progresses on transporting a photon with spatial information [25, 26], and on the storage and retrieval of an image in a hot atomic vapor [27, 28], we may expect the possible realization of a high dimensional quantum repeater in the future.

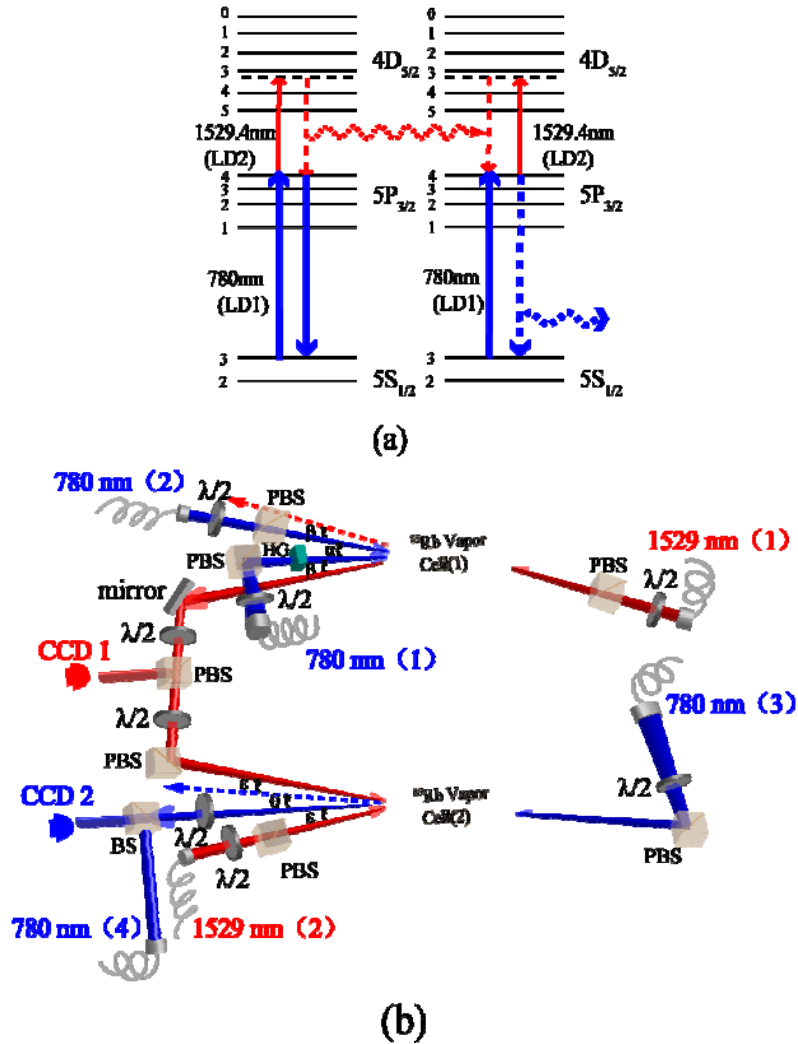

Fig.1. (Color online) (a) Energy level diagrams of the ladder-type configuration. (b) Schematic diagram of the experimental set-up for two sequential FWM processes. The red dashed line indicates the path of the 1529.4 nm signal generated from Rb vapor cell (1), and the blue dashed line that for the finally generated 780 nm signal from Rb vapor cell (2). HG: computer-generated hologram; PBS: polarization beam splitter; BS: beam splitter; $\lambda/2$ : half wave-plate.

To transfer an image imprinted on the light of wavelength 780 nm to the light in the telecom band and back again, two FWM processes occurring sequentially in two different hot atomic ensembles (two Rb vapor cells here) are used. For each ensemble, we choose a ladder-type atomic energy configuration with the energy level diagrams shown in Fig. 1(a); the left/right diagram corresponds to Rb vapor cell (1, 2). In the FWM process, a light of wavelength 780 nm from laser LD1, near the transition $5S_{1/2}(F=3)$ to $5P_{3/2}(F'=4)$, is incident on a homemade computer-generated hologram (HG). A HG with a fork structure at the centre can be used to change the order of the OAM of a beam. A Gaussian mode beam is in the Laguerre-Gaussian (LG) mode when it carries a well-defined OAM, and can be described by the $LG_{pl}$ mode, where $p+1$ is the number of the radial nodes, and $l$ is the number of the $2\pi$ phase variations along a closed path around the beam center. Here we consider the case of $p=0$, in which the $LG_{0l}$ mode carries the corresponding OAM of $l$. In our experiment, a first-order and a second-order HGs are used respectively. The $+/-$ order diffraction of the first-order (second-order) HG increases the OAM of the input beam by $\pm 1\hbar$ ($\pm 2\hbar$), when the dislocation of the HG overlaps with the beam centre. The superposition of the $LG_{00}$ mode and the $LG_{01}$ ($LG_{02}$) mode can be easily achieved by shifting the dislocation of the HG out of the beam centre a certain amount. After transmitting through the HG, the light of wavelength 780 nm from laser LD1 is divided into two beams: the zero-order diffraction wave is used as a reference, the other beam of being the +-order diffractive wave is incident on Rb vapor cell (1) as the input beam. FWM occurs if the input beam is combined with another light of 780 nm and light of 1529.4 nm near the transition $5P_{3/2}(F'=4)$ to $4D_{5/2}(F''=3)$ from a second laser LD2. These two beams are Gaussian modes. The generated light is directed towards Rb vapor cell (2) as input light, and transformed into output light of wavelength 780 nm through a second FWM, combined with light from laser LD1 (near transition $5S_{1/2}(F=3)$ to $5P_{3/2}(F'=4)$) and light from the laser LD2 (near resonance of $5P_{3/2}(F'=4)$ to $4D_{5/2}(F''=3)$). These two beams are also Gaussian modes. In both FWM processes, the frequencies and the momentums of the four waves satisfy the respective conservations of energy and momentum requirements. The maximal FWM efficiency (without OAM) achieved is about 54% [29].

The experimental setup is depicted in Fig. 1(b). A continuous wave (cw) laser of wavelength 780 nm (DL100, Toptica), frequency-stabilized to the $^{85}$Rb atomic transition of $5S_{1/2}(F=3) \rightarrow 5P_{3/2}(F'=4)$, is divided into four beams: labelled 780 nm (1-4). Two of the four beams, 780 nm (1) and 780 nm (2), with perpendicular linear polarizations are directed at the Rb vapor cell (1) with small incident angle of $\alpha = 0.7°$; the beam 780 nm (1) is incident on a HG, the zero-order diffraction wave from the HG is used as a reference, and the +-order diffraction wave is the input beam that is to be frequency converted. A second cw laser of wavelength 1529 nm (DL100, Prodesign, Toptica), frequency-stabilized to $^{85}$Rb atomic transition of $5P_{3/2}(F'=4) \rightarrow 4D_{5/2}(F''=3)$ by using a two-photon absorption technique, is divided into two beams: labelled 1529.4 nm (1-2). The beam 1529.4 nm (1) is directed at the Rb vapor cell (1), nearly anti-parallel to beam 780 nm (2). The angle $\beta$ between them is $\beta = 0.4°$. These two beams have orthogonal linear polarization directions. The generated FWM light of 1529.4 nm from the Rb vapor cell (1) is directed onto the Rb vapor cell (2) as input for subsequent frequency conversion. At the same time, the transferred image in the generated light is monitored by a charge-coupled device (CCD) camera CCD1 with the response in the telecom band. The beams 780 nm (3) and 1529.4 nm (2) have orthogonal linear polarizations and they are near counter-propagation through the Rb vapor cell (2). The small angle between them is about $\varepsilon = 0.5°$. The input beam of 1529.4 nm generated in the Rb vapor cell (1) has an angle of $2\varepsilon + \theta$ with the beam 1529.4 nm (2), and $\theta = 0.8°$. With further FWM in the Rb vapor cell (2), the generated FWM signal of 1529.4 nm in the Rb vapor cell (1) is converted back into a 780 nm signal that is recorded by CCD2 camera with the response in visible band. To check whether the coherence of the initial input light has been preserved, we perform two different check experiments. First, we use the superposition of the $LG_{00}$ mode and the $LG_{01}$ ($LG_{02}$) mode as an input light for subsequent frequency conversion. We check whether the superposition can be preserved through this sequential FWM process. Second, we let the zero-order diffraction wave of the HG (780 nm (4)) be a reference and the +-order diffraction wave be the input for sequential FWM process, these two beams are in coherent superposition. We combine the output light from the Rb vapor cell (2) through the sequential FWM process with the reference beam at a beam splitter to check on interference between these two beams. The experimental parameters are follows: the diameter of each of the beams 780 nm (1-4) was 1 mm, 3 mm, 3 mm and 1mm respectively; the power of the beams 780 nm (1-3) was 1.6 mW, 6.65 mW, 2.0 mW, respectively; the power of 780 nm (4) can be adjusted by a polarization beamsplitter and a half waveplate to match the output light from the Rb vapor cell (2) for high interference visibility. The beam diameter for both 1529.4 nm (1) and (2) was 2 mm, their power was 2.03 mW and 1.04 mW respectively. Both cells containing isotopically-pure $^{85}$Rb were 5 cm long. The Rb vapor cells (1) and (2) were heated to a temperature of $130\,°C$ and approximately $70\,°C$ respectively.

Firstly, we perform the experiment with the first-order HG. The +-order diffraction wave of beam 780 nm (1) with the doughnut-shape intensity distribution shown in Fig. 2(a) is incident on the Rb vapor cell (1) as the input for subsequent frequency conversion. After FWM in the Rb vapor cell (1), the intensity distribution of the generated FWM light recorded by CCD1 is shown in Fig. 2(b). The generated signal of 1529.4 nm is then directed at the Rb vapor cell (2), and subsequently converted back into the output signal of 780 nm by FWM. Here, the signal recorded by CCD2 is shown in Fig. 2(c). The plots of the intensity profiles of the image along two orthogonal directions are shown in Fig. 2(a), 2(b) and 2(c). Comparing these figures, we can see clearly that the main characters of the image have been preserved during the whole process. In order to give a more accurate and quantitative evaluation of the transfer process, we calculate the similarity $R$ between these images [30] and the calculated similarity between Fig. 2(a) and 2(b), between Fig. 2(b) and 2(c), and between Fig. 2(a) and 2(c) is 86.2%，81.9%，88.6% respectively. These further prove the successful image transfer.

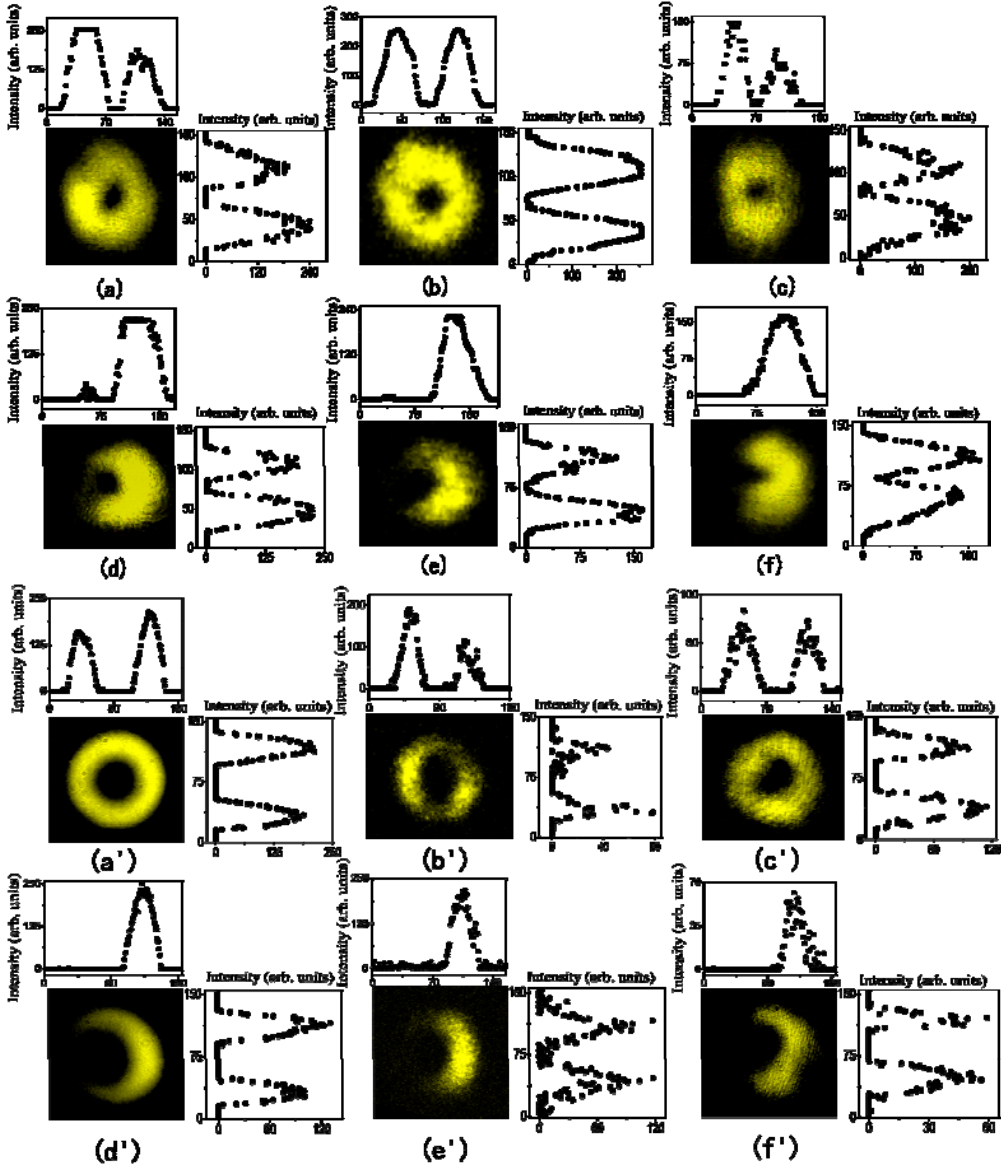

Fig. 2. (Color online) (a)-(f) are experimental results with the first-order HG, (a')-(f') obtained with the second-order HG. (a) and (a'): The image imprinted on the input light; (b) and (b'): the transferred image imprinted on the generated light of 1529.4 nm in the Rb vapor cell (1); (c) and (c'): the transferred image imprinted on the generated light of 780 nm through sequential FWM process in the Rb vapor cell (2). (d): the recorded input pattern caused by the superposition between the $LG_{00}$ mode and the $LG_{01}$ mode. (e): the transferred image imprinted on the generated light of 1529.4 nm through the first FWM and (f) the transferred image imprinted on the generated light of 780 nm through sequential FWM. (d')-(f') are experimental results obtained with the superposition between the $LG_{00}$ mode and the $LG_{02}$ mode. The dots in each figure are plots of intensity profiles along horizontal and vertical directions.

An important concern is whether the coherence of the input signal can be preserved during this sequential frequency conversion process. In the following, we perform two experiments to check it. In the first experiment, we use the superposition of the $LG_{00}$ mode and the $LG_{01}$ mode as an input for subsequent frequency conversion. The intensity distribution of this superposition is shown in Fig. 2(d). Experimentally we record the image of the superposition of the $LG_{00}$ mode and the $LG_{01}$ mode in the output beam from the Rb vapor cell (1) by CCD1, which is shown in Fig. 2e. Then the generated signal of 1529.4 nm from the Rb vapor cell (1) is directed at the Rb vapor cell (2), and subsequently converted back into an output signal of 780 nm by FWM. CCD2 camera recorded the signal and showed its intensity distribution in Fig. 2(f). The plots of the intensity profiles in two orthogonal directions are also shown in Fig. 2(e)-2(f). We also calculate the similarity between these figures [30], which is 96.9% between Fig. 2(d) and 2(e), 83.7% between Fig. 2(e) and 2(f), and 82.4% between Fig. 2(d) and 2(f). According to these results, we could conclude that the coherence between the $LG_{00}$ mode and the $LG_{01}$ mode is preserved during this sequential FWM process.

In the second experiment, the input beam is divided into two beams by a HG, the zero-order diffraction wave of the HG is used as the reference and +-order diffraction wave is the input for sequential FWM process. These two beams are in coherent superposition, which is proven by their interference shown in Fig. 3(a). After whole process, the reference is combined with the output signal from the Rb vapor cell (2), obtained from the sequential FWM process, at a beam splitter to record the interference pattern. Interference indicates that coherence between the twice-converted and reference beams is preserved. In the experiment, we observe the clear well-known interference pattern shown in Fig. 3(b). The similarity between Fig. 3(a) and 3(b) clearly shows that the coherence between the zero-order diffraction wave and the +-order diffraction wave is preserved during this sequential FWM process. The Fig. 3(a) and 3(b) are obtained when two interfering beams are overlapped completely along the propagation direction. If there is angle between them, then an interference pattern similar to the structure of the HG shown in Fig. 3(c) should be observed. In the experiment, we really observe such an interference pattern shown in Fig. 3(d): a fork structure of the HG.

After having completed the experiments with the first-order HG, we also perform the similar experiments with a second-order HG. The results are shown in Fig. 2(e')-2(f') and in Fig. 3(a')-3(d'). The calculated similarity [30] is 90.4% between Fig. 2(a') and 2(b'), 89.9% between Fig. 2(b') and 2(c'), 89.7% between Fig. 2(a') and 2(c'), and 91.4% between Fig. 2(d') and 2(e'), 86.1% between Fig. 2(e') and 2(f'), and 92.3% between Fig. 2(d') and 2(f'). These results also show the clear image transfer and the persistence of the coherence in the sequential FWM process.

Furthermore, we input a more complicate image obtained by the interference between the $LG_{00}$ mode and the $LG_{02}$ mode shown in Fig. 4(a) for further demonstration. Fig. 4(b) and 4(c) shows the image transfer through the first and the second FWM process respectively. The calculated similarity [30] between Fig. 4(a) and 4(b), between Fig. 4(b) and 4(c), and between Fig. 4(a) and 4(c) is 89.4%, 89.3%, 88.7% respectively. These results further demonstrate the success of the image transfer.

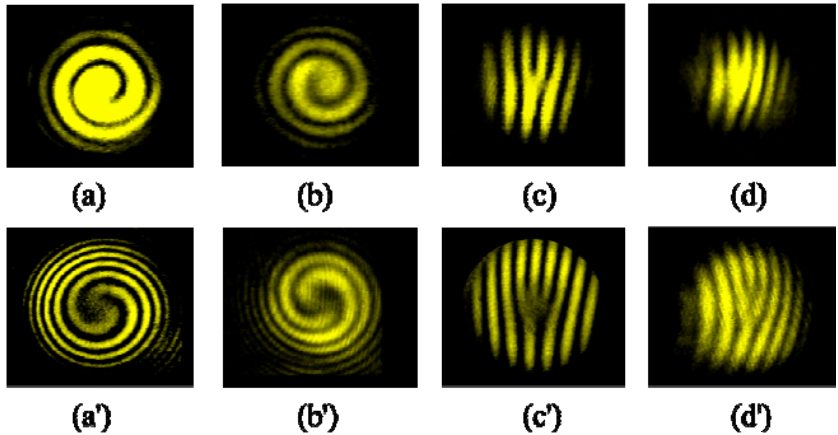

Fig. 3. (Color online) (a)-(d) are experimental results with the first-order HG, (a')-(d') obtained with the second-order HG. (a) and (a'): Interference pattern between the +-order diffraction wave of HG and the zero-order diffraction wave; (b) and (b'): the interference pattern between the +-order diffraction wave of HG through sequential FWM process with the zero-order diffraction wave as the reference. Two interfering beams are overlapped perfectly along the propagation

direction. (c), (c') and (d), (d') are observed patterns when there is angle between two interfering beams in the interference experiment similar to (a), (a') and (b), (b'). A fork structure of the HG can be clearly observed.

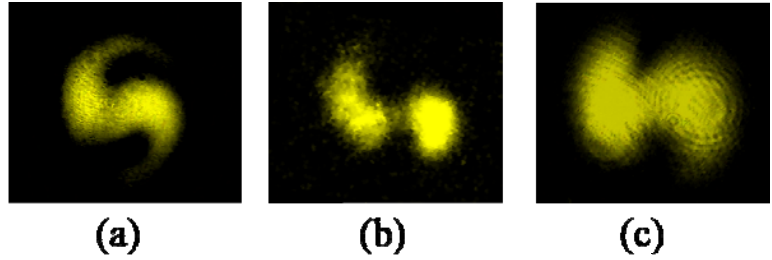

(a)  (b)  (c)

Fig.4. (Color online) (a): The image imprinted on the input light, which is obtained by the interference between the $LG_{00}$ mode and the $LG_{02}$ mode; (b): the transferred image imprinted on the generated light of 1529.4 nm through FWM in the Rb vapor cell (1) and (c): the transferred image imprinted on the generated light of 780 nm through sequential FWM process in the Rb vapor cell (2).

There exists the saturation effect of CCD in the experiment with the first-order HG because of strong input light, and it is avoided by using an attenuator to reduce the power of the input light when we perform the experiment with the second-order HG. Another thing we want to mention is that although our demonstration only involves three LG modes (00, 01 and 02) and the superposition of any two modes, we believe our results can be extended to a high number of modes. At present, it is difficult for us to produce the superposition of three or more modes by our HG, so we have not performed such a demonstration yet.

There are several factors which affect the quality of the transferred image, one is the diffractive effect of light during the propagation. The image imprinted on the generated FWM and propagating in the free space undergoes a paraxial diffraction spreading, becomes bold and eventually blurs. Additionally, the Rb media acts as a spatial filter due to the phase matching in FWM process, obstructing the spatial high frequencies upconversion, making the image further distorted. The geometry of our experimental set-up is another reason. The non-collinear configuration is used to reduce the noise from strong input lights in our experiment, it makes the FWMs along horizontal and vertical directions different, causing the image somehow asymmetric.

We also need to avoid the obvious self-focusing and self-defocusing effects [31], and the honeycomb pattern forming in the cell [32]. The appearing of these effects is strongly dependent on the power density of lasers (esp. of laser at 780 nm) and the atomic number density. Another factor is the interference among the lights reflected from the optical components, such as mirrors, cell faces, etc.

Although Lett *et. al.* reported on the generation of spatially multimode twin beams using FWM in a hot atomic vapor [33, 34] and Tabosa *et. al*. [35] reported on an experiment of the OAM transfer through a FWM process, their experiments showed significant differences to ours: a lambda-type atomic structure is used and all four beams in FWM are nearly of the same frequency in their experiments. Besides, they only realize a one-way frequency conversion process. In Ref. 8, two sequential FWM processes in two different cold atomic Rb ensembles are realized, but there is no image transfer. Besides, a complicated cooling and trapping experimental setup is used there.

In conclusion. We report on here the first demonstration of the image transfer accompanied by frequency conversion between the light of wavelength 780 nm and the light of wavelength 1530 nm by performing two sequential FWM processes in two different hot atomic Rb vapor cells. Moreover, we describe the confirming the persistence of the coherence of input light following application of the sequential FWM process experimentally. The results may have potential applications in optical communication, imaging, and quantum information fields.

**Acknowledgements**

DSD and BSS thank Dr. X-F Ren and Mr. L-L Wang very much for providing the HG and very useful advices, also thank Dr. F-Y Wang for useful discussion about the data analysis. This work was supported by the National Natural Science Foundation of China (Grant Nos. 10874171, 11174271), the National Fundamental Research Program of China (Grant No. 2011CB00200), the Innovation fund from CAS, Program for NCET.